# The Ballistic Pressure Wave Theory of Handgun Bullet Incapacitation

Michael Courtney, PhD
Ballistics Testing Group, P.O. Box 24, West Point, NY 10996
Michael_Courtney@alum.mit.edu

Amy Courtney, PhD
Department of Physics, United States Military Academy, West Point, NY 10996
Amy_Courtney@post.harvard.edu

**Abstract:**
This paper presents a summary of seven distinct chains of evidence, which, taken together, provide compelling support for the theory that a ballistic pressure wave radiating outward from the penetrating projectile can contribute to wounding and incapacitating effects of handgun bullets. These chains of evidence include the fluid percussion model of traumatic brain injury, observations of remote ballistic pressure wave injury in animal models, observations of rapid incapacitation highly correlated with pressure magnitude in animal models, epidemiological data from human shootings showing that the probability of incapacitation increases with peak pressure magnitude, case studies in humans showing remote pressure wave damage in the brain and spinal cord, and observations of blast waves causing remote brain injury.

## I. Introduction

Debates in terminal ballistics such as light-and-fast vs. slow-and-heavy debates are dramatic oversimplifications of the more scientific question of whether the wound channel (directly crushed tissue) is the only contributor to handgun bullet effectiveness or whether a more energy dependent parameter such as hydrostatic shock, the temporary stretch cavity, or ballistic pressure wave can also contribute.

These debates have been dominated by long-winded rhetoric and authoritative appeals rather than scientific data and analysis. Here, we summarize findings that support and quantify the pressure wave hypothesis:

*Other factors being equal, bullets producing larger pressure waves incapacitate more rapidly than bullets producing smaller pressure waves.*

The pressure wave hypothesis is supported by:
1) Pressure pulses inducing incapacitation and brain injury in laboratory animals [THG97, TLM05].
2) Ballistic pressure waves originating remotely from the brain causing measurable brain injury in pigs and dogs [SHS87, SHS88, SHS90a, SHS90b, WWZ04].
3) Experiments in animals showing the probability of rapid incapacitation increases with peak pressure wave magnitude [STR93, COC06c, COC06d, COC07a].
4) Epidemiological data showing that the probability of incapacitation increases with the peak pressure wave magnitude [MAS92, MAS96, COC06b].
5) Brain damage occurring without a penetrating brain injury in a human case study [THG96, COC07b].
6) Ballistic pressure waves causing spinal cord injuries in human case studies [STU98, SSW82, TAG57].
7) Blast waves causing brain injury without penetrating injury or blunt force trauma [MAY97, TAH98, CWJ01].

## II. What is the ballistic pressure wave?

The ballistic pressure wave is the force per unit area created by a ballistic impact that could be measured with a high-speed pressure transducer [COC06c]. The bullet slows down in tissue due to the retarding force the tissue applies to the bullet. In accordance with Newton's third law, the bullet exerts an equal and opposite force on the tissue.

The average pressure on the front of a bullet is the retarding force divided by the frontal area of the bullet. The pressure exerted by the medium on the bullet is equal to the pressure exerted by the bullet on the medium. Because the frontal area of a bullet is small, the pressure at the front of the bullet is large.

Once created, this pressure front travels outward in all directions in a viscous or visco-elastic medium such as soft tissue or ballistic gelatin. Propagating outward, the wave's decreasing magnitude results from the increasing total area the pressure wave covers.

To compare pressure waves produced by different loads, it is necessary to specify the distance from the center of the bullet path. For non-fragmenting JHP handgun bullets that expand reliably, the peak pressure wave magnitude (in PSI) on the edge of a 1" diameter cylinder concentric with the bullet path can be estimated as [COC06c]

$$p = \frac{5E}{d\pi},$$

where E is the kinetic energy (ft-lbs) of the bullet at impact, and d is the penetration depth (feet). The pressure wave is larger for fragmenting bullets [COC06b].

Wave magnitude falls off with increasing distance from the point of origin unless reflected by a boundary or confined to an internal structure. An internal pressure wave created in the thoracic cavity will be reflected multiple times by the sides of the cavity. Superposition of waves creates localized regions of high pressure by focusing the wave, just as concave mirrors focus light waves and concave surfaces focus sound waves.

Since pressure wave magnitude is inversely proportional to penetration depth, cutting penetration in half doubles the pressure, if kinetic energy is the same. However, the potential for increased incapacitation is limited, because the wave must be created inside soft tissue and close to major blood vessels or vital organs to have its effect. A bullet that barely penetrates the thoracic cavity has little effect. Incapacitation effects are reduced for penetration depths below 9.5 inches [COC06b].

### III. Fluid Percussion Model of TBI
The lateral fluid percussion model (LFP) of traumatic brain injury (TBI) is used to study mechanisms of traumatic brain injury [TLM05]. A brief pressure pulse is applied directly to the brain of a laboratory animal.

Both instantaneous incapacitation and neural damage can result [THG97]. Investigators have shown mild and moderate injury levels occur with pressure levels in the 15-30 PSI range. Pressure waves near 30 PSI caused immediate incapacitation in laboratory animals. It is widely believed that this model has direct application to cellular and mechanistic effects of TBI in humans [TLM05].

### IV. Animal models of remote brain injury
Suneson *et al.* [SHS87, SHS88, SHS90a, SHS90b] implanted high-frequency pressure transducers into the brains of pigs to measure pressures generated by missile impacts in the thigh. Transient pressure levels in the 18-45 PSI range were transmitted to the brain [SHS90a, fig 1].

Early tests [SHS87] observed apneic (non-breathing) periods after injury. Tissue analysis showed damage to brain-blood and blood-nerve barriers. Subsequent experiments [SHS90a, SHS90b] reported damage at the cellular level in the hippocampus and hypothalamus regions of the brain. This damage was apparently caused by pressure waves transmitted to the brain from the distant (0.5 m) point of origin.

Martin Fackler, former editor of the out of print **Wound Ballistics Review,** published negative reviews of these findings [FAC91a, FAC96a]. However, his critical reviews have been shown to contain exaggerations, logical fallacies, and scientific errors [COC06a].

For example, Fackler asserts that Suneson's findings are invalid because lithotripsy (ultrasonic kidney stone treatment) applies a large pressure wave without damaging tissue. However, pressure waves associated with lithotripsy have been shown to cause significant tissue injury [EWL98, LOS01, LKK03].

In addition, remote brain injury attributed to a ballistic pressure wave has also been found in a similar experiment in dogs. Independent scientists concluded [WWZ04]:

*These findings correspond well to the results of Suneson et al., and confirmed that distant effect exists in the central nervous system after a high-energy missile impact to an extremity. A high-frequency oscillating pressure wave with large amplitude and short duration was found in the brain after the extremity impact of a high-energy missile . . .*

These animal models provide compelling support for the pressure wave hypothesis.

### V. Animal models of incapacitation

The largest available data set quantifying handgun bullet incapacitation in animal test subjects (goats) [STR93] shows that average incapacitation time correlates strongly with ballistic pressure wave magnitude [COC06c].

A model for average incapacitation time in terms of peak pressure wave magnitude, p, is:

$$AIT(p) = 10s \times \sqrt{\frac{p_0}{p}},$$

where $p_0$ is an adjustable parameter that gives an average incapacitation time of 10 seconds. A least-squares fit gives $p_0$ = 482 PSI with a standard error of 1.64 s and a correlation coefficient of R = 0.91. A plot of AIT(p) is shown in Figure 1 along with the data.

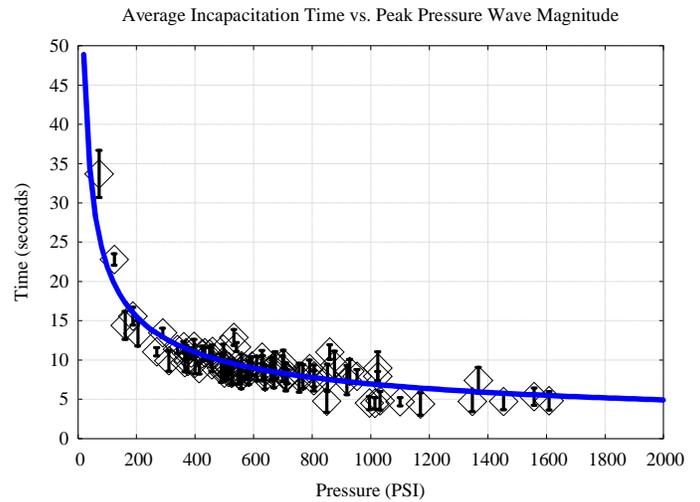

*Figure 1: A plot of average incapacitation time vs. pressure for the Strasbourg tests, along with the best-fit model.*

Fackler has also criticized this data set [FAC94a, FAC97a]. Without any eyewitness, documentary, or physical evidence showing fraud, he uses the opinion of a group of experts to assert that the report is fraudulent [FAC94a]:

*The FBI committee, which includes a half dozen of the world's most highly regarded gunshot-expert forensic pathologists, felt that the organization and wording of the document betrayed it as a hoax. Why else would experimental results be circulated anonymously?*

In a self-contradiction, these six "experts" are not named. The FBI committee remains anonymous while stating anonymity as the criterion used to determine that the Strasbourg report is a hoax![1]

Fackler's review also contains numerous fallacies leading a review to conclude [COC06a]:

*In the absence of support or direct contradiction from other experiments, the veracity of the Strasbourg tests should fairly*

---
[1] In the history of science, there are a number of examples of anonymous publication. Anonymity is not generally considered a conclusive indication of fraud [COC06a].

*be considered to be an open question. Neither the anonymity of the authors nor other criticisms offered are sufficient to consider the report fraudulent. Rather than lean too heavily on (possibly biased) expert opinions, the veracity of the report should be determined by the degree to which the reported results find support in other experimental findings.*

An experiment in deer finds quantitative agreement with the Strasbourg tests by using average incapacitation times in goats to predict average drop distances in deer [COC06d]. A separate experiment demonstrated incapacitation via a ballistic pressure wave in the absence of a wound channel [COC07a]. Analysis of the Strasbourg data also agrees with observations that a remote ballistic pressure wave reaches the brain and causes brain injury [SHS90a, SHS90b, WWZ04].

**VI. Analysis of Epidemiological Data**

Evan Marshall and Ed Sanow [MAS92, MAS96] compiled the largest available data set quantifying relative handgun bullet effectiveness in humans. Fackler again published critical articles [FAC94b, FAC97a, FAC99a, FAC99b]. In the article, "Undeniable Evidence," [FAC99b] he uses the bandwagon fallacy:

*From the outset, those with training in statistics, those schooled in the scientific method, those with experience in scientific research, and even those laymen who do their own thinking, have believed that the "one-shot stop" statistics published by Marshall were not collected as claimed, but simply made up – fabricated.*

A comprehensive review of these (and other) criticisms found [COC06a]:

*The published criticisms include unjustified ad hominem attacks and other rhetorical fallacies, gross exaggerations depending upon unjustified presuppositions, and valid concerns affecting the accuracy but not the validity of considering the OSS rating as a measure of relative handgun load effectiveness.*

Analysis of one-shot stop (OSS) data reveals contributions from both the ballistic pressure wave and the surface area, A, of the wound channel [COC06b]. Models for the probability of each incapacitation mechanism were combined by the rules of probability to derive a model for the total OSS rating.

The OSS model can be written as

$$OSS(p,A) = 100\% - 100\% \times \left[ \frac{1}{1+\left(\frac{p}{p_0}\right)^{\frac{3}{2}}} \times \frac{1}{1+\left(\frac{A}{A_0}\right)^{\frac{3}{2}}} \right],$$

where $p_0$ and $A_0$ are the best fit characteristic pressure wave magnitude and crush cavity surface area, respectively. A least squares fit finds $p_0$ = 339 PSI and $A_0$ = 31.3 sq. in. with a standard error = *5.6%* and a correlation coefficient of R = 0.939 [COC06b]. The model surface is graphed in Figure 2 with the OSS data set.[2]

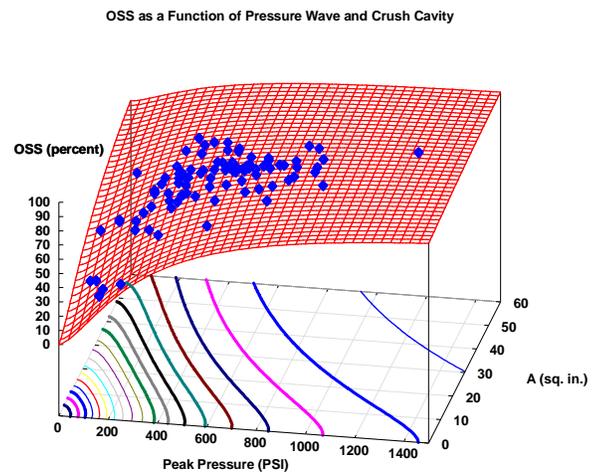

*Figure 2: The best-fit model of the OSS rating (red surface) plotted along with the OSS data (blue points).*

---

[2] This model compares favorably with Steve Fuller's "Best Fit" model [MAS96, Ch 28] that has a standard error of 4.79%. In contrast to Fuller's model, the pressure wave model uses only two adjustable parameters (with physical interpretations), two independent variables (related to incapacitation mechanisms), and gives the expected limiting behavior at small and large pressure wave and crush cavity sizes.

Contour lines at the bottom of the graph correspond to lines of constant OSS rating. These show that for a given OSS rating the permanent cavity size decreases as the peak pressure magnitude increases. Agreement between the pressure wave model and the OSS data set is compelling support for the pressure wave hypothesis.

## VII. Case Study of Remote Brain Injury

A World War II veteran received a bullet wound to the head but not directly impacting the brain [THG96]. The 7.62 mm Russian Tokarev pistol has a muzzle energy of approximately 430 ft-lbs. A penetration depth estimated at 8" yields a local pressure wave magnitude of 1027 PSI [COC07b].

The patient experienced acute epileptic symptoms that ceased after a few years under medication but returned nearly 50 years later. This is attributed to the "so-called hydrodynamic effect" (pressure wave) of the high velocity bullet causing an indirect trauma to the brain [THG96].

## VIII. Case Studies of Remote Spinal Cord Injuries

The brain is not the only organ subject to remote pressure wave effects. In a study of handgun injury, Sturtevant found that pressure waves from a bullet impact in the torso can reach the spine. Moreover, a focusing effect from concave surfaces can concentrate the pressure wave on the spinal cord producing significant injury [STU98]. This is consistent with other case studies in humans showing remote spinal cord injuries from ballistic impacts [TAG57, SSW82].

## IX. Blast injury

Blast injury can be caused by an externally imposed pressure wave of an explosion without penetrating injury or blunt force trauma [MAY97, TAH98, CWJ01]. The internal pressure created by the interaction of bullet and tissue can be larger than the external pressures associated with blast injury. The interaction of tissue with a pressure wave depends only on the characteristics of the wave. Tissue damage will be similar for similar wave characteristics.

Dr. Ibolja Cernak, a leading researcher in blast wave injury at the Applied Physics Laboratory at Johns Hopkins University, hypothesized, "alterations in brain function following blast exposure are induced by kinetic energy transfer of blast overpressure via great blood vessels in abdomen and thorax to the central nervous system" [CER05]. This hypothesis is supported by observations of neural effects in the brain from localized blast exposure focused on the lungs in animal experiments.

Consider the experiment where a pressure wave creates incapacitation when a bullet is fired into water close to a test animal [COC07a]. The externally applied pressure wave is not much different from an explosion that transfers a similar amount of energy in a similar amount of time. If externally applied blast pressure waves can cause traumatic brain injury, it stands to reason that internally applied ballistic pressure waves can also.

## X. An Emerging Theory

Individually, the various results discussed above each suggest some level of support for the pressure wave hypothesis. Taken together, these observations and experiments provide separate chains of support for the emerging ballistic pressure wave theory.

Links between traumatic brain injury and the ballistic pressure wave suggest that brain injury begins to be possible for pressures above 500 PSI applied inside the chest and brain injury becomes probable with 1000 PSI [COC07b]. These are probably reasonable estimates for the pressures associated with incapacitation.

Figure 3 shows the probability of rapid incapacitation for a given pressure wave applied in the chest [COC06c]. The human model (in red) is based on the goat model (in blue) and the idea that humans are more susceptible than goats and deer to larger pressure waves.

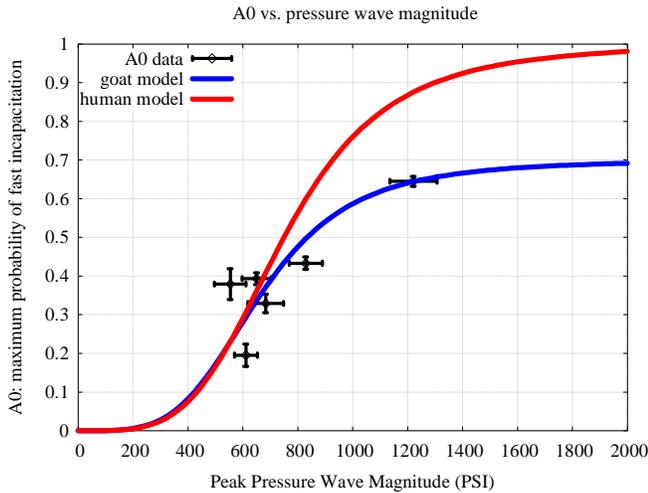

Figure 3: Probability of rapid incapacitation for a given pressure wave applied to the chest [COC06c].

Humans are almost always immediately incapacitated by rapidly expanding or fragmenting .308 bullets to the chest [MAS96], whereas a significant fraction of deer remain on their feet for 5 seconds or so until they collapse from loss of blood. In addition, ruminants (such as deer and goats) are more resistant to brain injury than humans because their pre-mating dominance rituals involve head butting [SHA02].

There is no magic bullet, but loads that can produce over 1000 PSI in the chest tend to be more effective. Figure 4 shows pressure wave magnitudes as a function of energy for bullets penetrating 10, 12, and 14 inches and retaining 100% and 55% of their initial mass.

Fragmenting bullets produce greater pressure magnitude than non-fragmenting bullets. Bullets penetrating 10-12" produce more pressure than bullets penetrating 14" or more. For example, a bullet which does not fragment and penetrates 14" needs over 700 ft-lbs of energy to produce 1000 PSI. In contrast, a bullet which fragments and penetrates 12" needs just under 450 ft-lbs of energy to produce 1000 PSI.

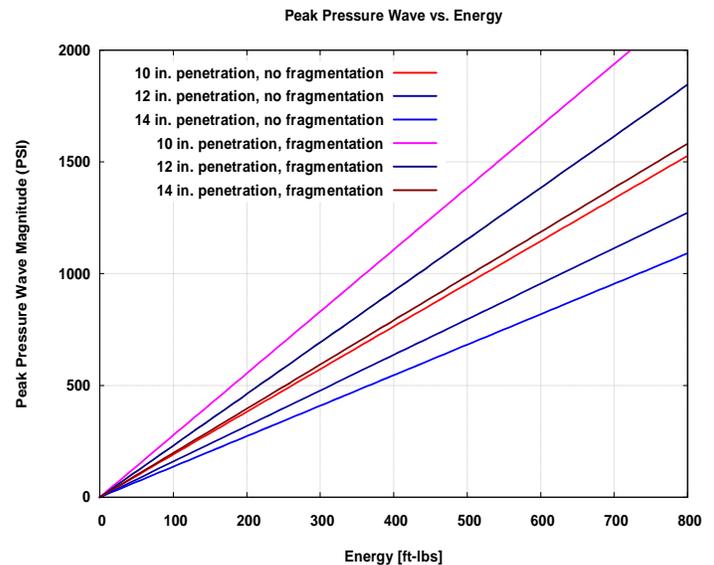

Figure 4: Peak pressure wave vs energy for different penetration depths. Graphs for fragmenting bullets represent bullets retaining 55% of their initial mass.

**XI. Conclusion and Limits of Interpretation**

The scientific foundation for ballistic pressure wave contributions to bullet effectiveness does not suggest that the pressure wave is the only contributor to incapacitation. The permanent cavity also plays an important role. The relative importance of these mechanisms is a matter for additional research.

One should not be overly impressed by the propensity for shallow penetrating loads to produce larger pressure waves. Selection criteria should first determine the required penetration depth for the given risk assessment and application, and only use pressure wave magnitude as a selection criterion for loads meeting minimum penetration requirements.

Reliable expansion, penetration, feeding, and functioning are all important aspects of load testing and selection. We do not advocate abandoning long-held aspects of the load testing and selection process, but it seems prudent to consider the pressure wave magnitude along with other factors.

**References:**
[CER05] Cernak I, Blast (Explosion)-Induced Neurotrauma: A Myth Becomes Reality, Restorative Neurology and Neuroscience, 23:139-140, 2005.

[COC06a] Courtney M, Courtney A: Review of criticisms of ballistic pressure wave experiments, the Strasbourg goat tests, and the Marshall and Sanow data. http://arxiv.org/ftp/physics/papers/0701/0701268.pdf

[COC06b] Courtney M, Courtney A: Relative incapacitation contributions of pressure wave and wound channel in the Marshall and Sanow data set. http://arxiv.org/ftp/physics/papers/0701/0701266.pdf

[COC06c] Courtney M, Courtney A: Ballistic pressure wave contributions to rapid incapacitation in the Strasbourg goat tests. http://arxiv.org/ftp/physics/papers/0701/0701267.pdf

[COC06d] Courtney M, Courtney A: A method for testing handgun bullets in deer. http://arxiv.org/ftp/physics/papers/0702/0702107.pdf

[COC07a] Courtney M, Courtney A: Experimental Observations of Incapacitation via Ballistic Pressure Wave without a Wound Channel, www.ballisticstestinggroup.org/lotor.pdf

[COC07b] Courtney A, Courtney M. Links between traumatic brain injury and ballistic pressure waves originating in the thoracic cavity and extremities. Brain Inj 2007;21:657-662. Pre-print: www.ballisticstestinggroup.org/tbipwave.pdf

[CWJ01] Cernak I, Wang Z, Jiang J, Bian X, Savic J: Ultrastructural and functional characteristics of blast injury-induced neurotrauma. Journal of Trauma 2001;50(4):695-706

[EWL98] Evan AP, Willis LR Lingeman JE, McAteer JA: Editorial: Renal Trauma and the Risk of Long-Term Complications in Shock Wave Lithotripsy. Nephron 78(1):1-8, 1998.

[FAC91a] Fackler ML: Literature Review and Comment. Wound Ballistics Review Winter 1991: pp38-40.

[FAC94a] Fackler ML: The 'Strasbourg Tests:' Another Gunwriter/Bullet Salesman Fraud? Wound Ballistics Review 1(4):10-11; 1994.

[FAC94b] Fackler ML: Marshall-Sanow Can't Beat the Long Odds: Wound Wizards Tally Too Good to be True. Soldier of Fortune January:64-65; 1994.

[FAC96a] Fackler ML: Gunshot Wound Review. Annals of Emergency Medicine 28(1): 194-203; 1996.

[FAC97a] Fackler ML: Book Review: Street Stoppers – The Latest Handgun Stopping Power Street Results. Wound Ballistics Review 3(1):26-31; 1997.

[FAC99a] Fackler ML: Editorial. Wound Ballistics Review 4(2):15-16; 1999.

[FAC99b] Fackler ML: Undeniable Evidence. Wound Ballistics Review 4(2):15-16; 1999.

[LKK03] Lingeman JE, Kim SC, Keo RL, McAteer JA, Evan AP: Shockwave Lithotripsy: Anecdotes and Insights. Journal of Endourology 17(9):687-693; 2003.

[LOS01] Lokhandwalla M, Sturtevant B: Mechanical Haemolysis in Shock Wave Lithotripsy (SWL): I. Analysis of Cell Deformation due to SWL Flow-Fields." Physics in Medicine & Biology 46(2):413-437; 2001.

[MAS92] Marshall EP and Sanow EJ: Handgun Stopping Power: The Definitive Study. Paladin Press, Boulder, CO, 1992.

[MAS96] Marshall EP and Sanow EJ: Street Stoppers. Paladin Press, Boulder, CO, 1996.

[MAY97] Mayorga MA: The pathology of primary blast overpressure injury. Toxicology 1997;121:17-28

[SHS87] Suneson A, Hansson HA, Seeman T: Peripheral High-Energy Missile Hits Cause Pressure Changes and Damage to the Nervous System: Experimental Studies on Pigs. The Journal of Trauma. 27(7):782-789; 1987.

[SHS88] Suneson A, Hansson HA, Seeman T: Central and Peripheral Nervous Damage Following High-Energy Missile Wounds in the Thigh. The Journal of Trauma. 28(1 Supplement):S197-S203; January 1988.

[SHS90a] Suneson A, Hansson HA, Seeman T: Pressure Wave Injuries to the Nervous System Caused by High Energy Missile Extremity Impact: Part I. Local and Distant Effects on the Peripheral Nervous System. A Light and Electron Microscopic Study on Pigs. The Journal of Trauma. 30(3):281-294; 1990.

[SHS90b] Suneson A, Hansson HA, Seeman T: Pressure Wave Injuries to the Nervous System Caused by High Energy Missile extremity Impact: Part II. Distant Effects on the Central Nervous System. A Light and Electron Microscopic Study on Pigs. The Journal of Trauma. 30(3):295-306; 1990.

[SSW82] Saxon M, Snyder HA, Washington HA, Atypical Brown-Sequard syndrome following gunshot wound to the face. J Oral and Maxillofacial Surgery 1982;40:299-302.

[STR93] The Strasbourg Tests, presented at the 1993 ASLET International Training Conference, Reno, Nevada.

[SHA02] Shaw NA: The Neurophysiology of Concussion. Progress in Neurobiology 67:281-344; 2002.

[STU98]  Sturtevant B, Shock Wave Effects in Biomechanics, Sadhana, 23: 579-596, 1998.

[TAG57]  Taylor RG, Gleave JRW. Incomplete Spinal Cord Injuries. J Bone and Joint Surgery 1957;B39:438-450.

[TAH98] Trudeau DL, Anderson J, Hansen LM, Shagalov DN, Schmoller J, Nugent S, Barton S: Findings of mild traumatic brain injury in combat veterans with PTSD and Mayorga MA: The pathology of primary blast overpressure injury. Toxicology 1997;121:17-28.

[THG96]  Treib J, Haass A, Grauer MT: High-velocity bullet causing indirect trauma to the brain and symptomatic epilepsy. Military Medicine 1996;161(1):61-64.

[THG97]  Toth Z, Hollrigel G, Gorcs T, and Soltesz I: Instantaneous Perturbation of Dentate Interneuronal Networks by a Pressure Wave Transient Delivered to the Neocortex.  The Journal of Neuroscience 17(7);8106-8117; 1997.

[TLM05]  Thompson HJ, Lifshitz J, Marklund N, Grady MS, Graham DI, Hovda DA, McIntosh TK: Lateral Fluid Percussion Brain Injury: A 15-Year Review and Evaluation. Journal of Neurotrauma 22(1):42-75; 2005.

[WWZ04]  Wang Q, Wang Z, Zhu P, Jiang J: Alterations of the Myelin Basic Protein and Ultrastructure in the Limbic System and the Early Stage of Trauma-Related Stress Disorder in Dogs. The Journal of Trauma. 56(3):604-610; 2004.

**About the Authors**

*Amy Courtney* currently serves on the faculty of the United States Military Academy at West Point.  She earned a MS in Biomedical Engineering from Harvard University and a PhD in Medical Engineering and Medical Physics from a joint Harvard/MIT program.   She has taught Anatomy and Physiology as well as Physics.  She has served as a research scientist at the Cleveland Clinic and Western Carolina University, as well as on the Biomedical Engineering faculty of The Ohio State University.  Amy_Courtney@post.harvard.edu

*Michael Courtney* earned a PhD in experimental Physics from the Massachusetts Institute of Technology.  He has served as the Director of the Forensic Science Program at Western Carolina University and also been a Physics Professor, teaching Physics, Statistics, and Forensic Science.   Michael and his wife, Amy, founded the Ballistics Testing Group in 2001 to study incapacitation ballistics and the reconstruction of shooting events. Michael_Courtney@alum.mit.edu